\documentclass[12pt,thmsa]{article}
\usepackage{sw20lart}

\input{tcilatex}
\begin{document}

\title{Dark energy as space-time curvature induced by quantum vacuum fluctuations }
\author{Emilio Santos \\
Departamento de F\'{i}sica. Universidad de Cantabria. Santander. Spain}
\maketitle

\begin{abstract}
It is shown that quantum vacuum fluctuations give rise to a curvature of
space-time equivalent to a cosmological constant, that is a homogeneous
energy density $\rho $ and pressure $p$ fulfilling -$p$\ = $\rho >0.$ The
fact that the fluctuations produce curvature, even if the vacuum expectation
of the energy vanishes, is a consequence of the non-linear character of the
Einstein equation. A calculation is made, involving plausible hypotheses
within quantized gravity, which establishes a relation between the two-point
correlation of the vacuum fluctuations and the space-time curvature.
Arguments are given which suggest that the density $\rho $ might be of order
the ``dark energy'' density currently assumed to explain the observed
accelerated expansion of the universe.

\textbf{Keywords }Dark energy. Vacuum fluctuations. Quantum gravity

\end{abstract}

\section{ Introduction}

Recent astronomical observations, in particular the study of type Ia
supernovae, anisotropies in the cosmic background radiation and matter power
spectra inferred from large galaxy surveys, have improved our knowledge of
the universe giving rise to a precision cosmology. The new data are
compatible with the universe having a Friedmann - Robertson - Walker metric
with flat spatial slices\cite{Sahni} of the form 
\begin{equation}
ds^{2}=-dt^{2}+a(t)^{2}\left( dr^{2}+r^{2}d\Omega \right) .  \label{2.0}
\end{equation}
The time-dependent parameter $a(t)$ is related, at present time $t_{0}$, to
the measurable Hubble constant, $H_{0}$, and decceleration parameter, $q_{0}$%
, via 
\begin{equation}
\left[ \frac{\stackrel{.}{a}}{a}\right] _{t_{0}}=H_{0},\text{ }\left[ \frac{%
\ddot{a}}{a}\right] _{t_{0}}=-H_{0}^{2}q_{0}.  \label{2.1}
\end{equation}
The observations also provide information about the evolution of the
parameter $a(t)$ and the distribution of matter in the past.

The available knowledge is summarized in the $\Lambda CDM$ model. In it
baryonic matter density, $\rho _{B},$ represents about 4.6\% of the matter
content of the universe while two hypothetical ingredients named cold dark
matter ($CDM$) and dark energy ($DE$) contribute with densities $\rho
_{DM}\sim $ 23\%, and $\rho _{DE}\sim $ 73\% respectively\cite{Hinshaw}. The
said densities are related to the metric eq.$\left( \ref{2.0}\right) $ via
the Friedmann equations, derived from general relativity, giving the
following relations\cite{Rich} 
\begin{eqnarray}
\left[ \frac{\stackrel{.}{a}}{a}\right] ^{2} &=&\frac{8\pi G}{3}\left( \rho
_{B}\left( t\right) +\rho _{DM}\left( t\right) +\rho _{DE}\right) , 
\nonumber \\
\frac{\ddot{a}}{a} &=&\frac{8\pi G}{3}\left( \frac{1}{2}\left[ \rho
_{B}\left( t\right) +\rho _{DM}\left( t\right) \right] -\rho _{DE}\right) ,\;
\label{2.00}
\end{eqnarray}
where I neglect small effects of radiation and matter pressure. The baryonic
density $\rho _{B}$ is well known from the measured abundances of light
chemical elements, which allows calculating $\rho _{DE}$ and $\rho _{DM}$
from the empirical quantities $H_{0}$ and $q_{0}$ via comparison of eqs.$%
\left( \ref{2.00}\right) $ and $\left( \ref{2.1}\right) $. The values
obtained by this method agree with data from other observations. For
instance cold dark matter, in an amount compatible with $\rho _{DM},$ is
needed in order to explain the observed (almost flat) rotation curves in
galaxies. However the nature of dark matter and dark energy remain open
problems.

Usually dark matter is assumed to derive from exotic particles, not yet
discovered, and dark energy is identified with a cosmological constant, $%
\Lambda $. In any case the $\Lambda CDM$ model rest upon the assumption that
general relativity (GR) is indeed the correct theory of gravity. However it
is conceivable that both cosmic speed up and dark matter represent signals
of a breakdown of GR. For instance we might consider the possibility that
the Hilbert - Einstein Lagrangian, linear in the Ricci scalar $R$, should be
generalized to become a function $f(R)$. This is the underlying philosophy
of what is referred to as $f(R)$ gravity theory\cite{Faraoni}. Indeed the
cosmological constant corresponds to a particular choice of $f(R)$ where a
constant $\Lambda $ is added to the Ricci scalar $R$, although this does not
give any hint about the value of $\rho _{DE}.$ The theory of $f(R)$ gravity
provides sufficient freedom to accomodate also dark matter. For instance it
allows good fits to the rotation curves in galaxies\cite{JCAP}, which
therefore might be explained as a curvature effect. Actually $f(R)$ gravity
may be further generalized by including other scalars, like $R_{\mu \nu
}R^{\mu \nu },$ in the Hilbert - Einstein Lagrangian\cite{ES}. In this paper
I shall not deal with dark matter, but only with dark energy, so that I will
take the density $\rho _{DM}$ as an empirical datum, without any discussion
about its possible nature.

Many proposals have been made for the origin of dark energy (for a review
see Copeland\cite{Copeland}). As said above the most popular one is to
identify it with a cosmological constant or, what is equivalent in practice,
to assume that it derives from the quantum vacuum. Indeed the term $\rho
_{DE}$ in eqs.$\left( \ref{2.00}\right) $ might be interpreted as coming
from a vacuum energy whose pressure fulfils $p_{DE}$ $=-\rho _{DE},$ an
equality appropriate for the vacuum (in Minkowski space, or when the
space-time curvature is small) because it is invariant under Lorentz
transformations. A problem appears however when one attempts to estimate the
value of $\rho _{DE}.$ For instance if the dark energy is due to the
interplay between quantum mechanics and gravity, it may seem that it should
be either strictly zero or of order Planck\'{}s density, that is 
\begin{equation}
\rho _{DE}\sim \frac{c^{5}}{G^{2}
\rlap{\protect\rule[1.1ex]{.325em}{.1ex}}h%
}\simeq 10^{97}\text{ kg/m}^{3}.  \label{001}
\end{equation}
In sharp contrast the observations lead to 
\begin{equation}
\rho _{DE}\simeq 10^{-26}\text{ kg/m}^{3},  \label{01}
\end{equation}
a value departing from eq.$\left( \ref{001}\right) $ by about 123 orders of
magnitude. This strong disagreement gives rise to the ``cosmological
constant problem''\cite{Weinberg}.

In this paper I will explain why the value of $\rho _{DE}$ can be much
smaller than eq.$\left( \ref{001}\right) $ even if the cosmological constant
really derives from the quantum vacuum and I shall do that without departing
from standard general relativity. In fact I shall prove, modulo a few
plausible assumptions within quantum gravity, that vacuum fluctuations give
rise to a curvature of space-time fully equivalent to the one produced by a
cosmological constant, even if the vacuum expectation of the density of
quantum fields is strictly zero. The argument is as follows.

For a small enough region of the universe around us, but large in comparison
with typical distances between galaxies, the space-time metric given by eqs.$%
\left( \ref{2.0}\right) $ may be rewritten, near present time, using new
coordinates as follows\cite{Rich} 
\begin{equation}
ds^{2}\simeq g_{rr}dr^{2}+r^{2}d\Omega ^{2}-g_{tt}dt^{2},g_{rr}=\left[
1+\left[ \frac{\stackrel{.}{a}}{a}\right] _{0}^{2}r^{2}\right]
,g_{tt}=\left[ 1+\left[ \frac{\ddot{a}}{a}\right] _{0}r^{2}\right]
\label{2.2}
\end{equation}
where terms of order $O\left( r^{4}\right) $ have been neglected and it is
assumed that the (slow) change of the metric coefficients with time may be
ignored. This metric is Minkowskian for small $r,$ which makes the
calculations more simple than using eq.$\left( \ref{2.0}\right) $. In this
paper I will calculate the coefficients $g_{rr}$ and $g_{tt}$ of eq.$\left( 
\ref{2.2}\right) $ as coming from the combined action of cold matter, having
homogeneous density $\rho _{B}+\rho _{DM}$ (at the large scale), plus the
effect of the vacuum fluctuations. In my approach \textit{the Friedmann eqs.}%
$\left( \ref{2.00}\right) $ \textit{are not valid }because they were derived
under the assumption that the space-time curvature, resulting in the metric
eq.$\left( \ref{2.2}\right) ,$ comes from a mixture of three fluids with
total density $\rho _{B}+\rho _{DM}+\rho _{DE}$ and total pressure $-\rho
_{DE}.$ Here I will assume only two fluid with total density $\rho _{B}+\rho
_{DM}$ and negligible pressure. The calculation leads to the following
relations 
\begin{eqnarray}
g_{rr} &=&1+\left[ \frac{8\pi G}{3}\left( \rho _{B}\left( t\right) +\rho
_{DM}\left( t\right) \right) +\Lambda _{fluct}\right] r^{2}+O\left(
r^{4}\right) ,  \nonumber \\
g_{tt} &=&1+\left[ \frac{8\pi G}{3}\left( \frac{1}{2}\left[ \rho _{B}\left(
t\right) +\rho _{DM}\left( t\right) \right] \right) -\Lambda _{fluct}\right]
r^{2}+O\left( r^{4}\right) ,\;  \label{rofluct}
\end{eqnarray}
where $\Lambda _{fluct}$ is a constant parameter with dimentions of inverse
length squared. It is explicitly calculated from the two-point correlation
of the vacuum fluctuations (see below). The net result is that the
fluctuations produce the same effect on the space-time curvature as a
cosmological constant. Thus eq.$\left( \ref{rofluct}\right) $ may be written
in a form similar to Friedman\'{}s eqs.$\left( \ref{2.00}\right) $ provided
we define a new quantity, $\rho _{DE},$ with dimensions of density as
follows 
\begin{equation}
\rho _{DE}\equiv \frac{3}{8\pi G}\Lambda _{fluct}.  \label{rode}
\end{equation}
But $\rho _{DE}$ \textit{is not any actual density, but a parameter taking
into account the effect of the quantum vacuum fluctuations on space-time
curvature. }My derivation does not provide a precise value of $\rho _{DE}$
but it strongly suggests that it is far smaller than eq.$\left( \ref{001}%
\right) .$ The difficulty for getting the value is that the two-point
correlation of the vacuum fluctuations is not known. If future calculations
along this line provide a value of $\rho _{DE}$ in agreement with eq.$\left( 
\ref{01}\right) $ then the universe speed up would be fully explained as due
to the quantum vacuum fluctuations.

\section{The two-point correlation function of vacuum fluctuations}

The starting point of the work is an idea of Zeldovich\cite{Zel}, who
proposed the relations

\begin{equation}
\rho _{DE}c^{2}\sim G\frac{m^{6}c^{4}}{
\rlap{\protect\rule[1.1ex]{.325em}{.1ex}}h%
^{4}}=\frac{Gm^{2}}{\lambda }\times \frac{1}{\lambda ^{3}},\lambda \equiv 
\frac{
\rlap{\protect\rule[1.1ex]{.325em}{.1ex}}h%
}{mc}  \label{02}
\end{equation}
where $m$ is a typical mass of elementary particles, eq.$\left( \ref{01}%
\right) $ being obtained if the mass $m$ is 
\[
m\sim 7.6\times 10^{-29}\text{ kg,} 
\]
that is about one third the pion mass. In my opinion the Zeldovich\'{}s
relation between cosmology and particle physics suggests that dark energy
density eq.$\left( \ref{01}\right) $ does not correspond to the mean vacuum
energy, which is likely zero, but to (small) gravity effects associated to
the quantum vacuum fluctuations.

The second eq.$\left( \ref{02}\right) $ suggests that $\rho _{DE}c^{2}$ may
have the magnitude of the gravitational energy of quantum vacuum
fluctuations. This may be seen more explicitly using a semiclassical
Newtonian theory of gravity, that is taking matter as quantized but the
gravitational field as classical. Thus we may calculate the vacuum
expectation of the gravitational energy associated to a sphere of radius $R$%
, that is 
\begin{equation}
E=-G\int_{\left| \mathbf{r}_{2}\right| \leq R}d^{3}\mathbf{r}%
_{1}\int_{\left| \mathbf{r}_{2}\right| \leq R}d^{3}\mathbf{r}_{2}\frac{\frac{%
1}{2}\left\langle vac\left| \stackrel{\wedge }{\rho }\left( \mathbf{r}%
_{1},t\right) \stackrel{\wedge }{\rho }\left( \mathbf{r}_{2},t\right) +%
\stackrel{\wedge }{\rho }\left( \mathbf{r}_{2},t\right) \stackrel{\wedge }{%
\rho }\left( \mathbf{r}_{1},t\right) \right| vac\right\rangle }{\left| 
\mathbf{r}_{1}-\mathbf{r}_{2}\right| },  \label{2c}
\end{equation}
This calculation rests upon the assumption that it is possible to define an
energy density operator, $\stackrel{\wedge }{\rho }\left( \mathbf{r,}%
t\right) ,$ of the quantum fields and that the vacuum expectation of that
energy is zero but the expectation of the square is not zero. Actually if
the expectation of $\stackrel{\wedge }{\rho }^{2}$was also zero there would
be no quantum fluctuations at all. That is we must assume 
\begin{equation}
\left\langle vac\left| \stackrel{\wedge }{\rho }\left( \mathbf{r,}t\right)
\right| vac\right\rangle =0,\;\left\langle vac\left| \stackrel{\wedge }{\rho 
}^{2}\right| vac\right\rangle \neq 0.  \label{2}
\end{equation}
This being the case, by continuity we expect a two-point correlation
function which may depend only on the distance $\left| \mathbf{r}_{2}-%
\mathbf{r}_{1}\right| $ for equal times in the non-relativistic approach
leading to eq.$\left( \ref{2c}\right) .$ In a relativistic theory the
correlation should depend on the interval, $s$, this being the only
invariant in Minkowski space. The operators $\stackrel{\wedge }{\rho }\left( 
\mathbf{r}_{1},t_{1}\right) $ and $\stackrel{\wedge }{\rho }\left( \mathbf{r}%
_{2},t_{2}\right) $ may not commute and it is plausible to define the
two-point correlation with the operators in symmetrical order, which
guarantees that the correlation, $C(s),$ is real, without an imaginary part.
That is 
\begin{eqnarray}
C\left( s\right) &=&\frac{1}{2}\left\langle vac\left| \stackrel{\wedge }{%
\rho }\left( \mathbf{r}_{1},t_{1}\right) \stackrel{\wedge }{\rho }\left( 
\mathbf{r}_{2},t_{2}\right) +\stackrel{\wedge }{\rho }\left( \mathbf{r}%
_{2},t_{2}\right) \stackrel{\wedge }{\rho }\left( \mathbf{r}%
_{1},t_{1}\right) \right| vac\right\rangle ,  \label{3a} \\
s^{2} &=&\left( \mathbf{r}_{1}-\mathbf{r}_{1}\right) ^{2}-\left(
t_{1}-t_{2}\right) ^{2},  \nonumber
\end{eqnarray}
with units $c=1$ which I shall use from now on. With this definition the
correlation might depend on whether the interval $s$ is space-like or
time-like, in the latter case $s$ being imaginary.

Taking eq.$\left( \ref{3a}\right) $into account and considering equal times, 
$t_{1}=t_{2},$ eq.$\left( \ref{2c}\right) $ leads to 
\begin{eqnarray}
E &=&-G\int_{\left| \mathbf{r}_{2}\right| \leq R}d^{3}\mathbf{r}%
_{1}\int_{\left| \mathbf{r}_{2}\right| \leq R}d^{3}\mathbf{r}_{2}\frac{%
C\left( \left| \mathbf{r}_{1}-\mathbf{r}_{2}\right| \right) }{\left| \mathbf{%
r}_{1}-\mathbf{r}_{2}\right| }  \label{eg} \\
&\simeq &-G\int_{\left| \mathbf{r}_{2}\right| \leq R}d^{3}\mathbf{r}%
_{1}\int_{0}^{\infty }4\pi r^{2}\frac{C\left( r\right) }{r}dr=-4\pi
GV\int_{0}^{\infty }C\left( r\right) rdr,  \nonumber
\end{eqnarray}
where $r$ stands for $\left| \mathbf{r}_{1}-\mathbf{r}_{2}\right| $ and the
integral in $r$ has been extended to $\infty $ because we assume that the
radius $R$ is much larger than the range of the correlation function $%
C\left( r\right) .$ The result shows that, in Newtonian gravity, the
existence of fluctuations necessarily implies a gravitational energy
associated to them, with density 
\begin{equation}
\rho _{grav}=-4\pi G\int_{0}^{\infty }C\left( r\right) rdr.  \label{3c}
\end{equation}
The gravitational energy appears in spite of the fact that the vacuum
expectation of the matter density operator vanishes everywhere, see eq.$%
\left( \ref{2}\right) .$ The fact that semiclassical Newtonian theory
predicts that vacuum fluctuations give rise to a gravitational energy, even
if the mean energy density of the vacuum is strictly zero, shows that in
semiclassical general relativity the vacuum fluctuations will produce
space-time curvature. Indeed in general relativity the concepts of
gravitational force and gravitational energy lose their meaning and the
relevant concept is the curvature of space-time.

The aim of this paper is to present a calculation using general relativity.
Thus the goal of our calculation will be to find the curvature of space-time
induced by the quantum vacuum fluctuations. The calculation develops an idea
put forward elsewhere\cite{Santos}. Before presenting the calculation I
shall discuss further the subject of the two-point correlations of vacuum
energy densities. This is necessary because quantum vacuum fluctuations
might be seen as artifacts of the quantum formalism, without real physical
implications, in view that in most cases they may be eliminated by using
normal ordering of the creation and annihilation operators.

The two-point function $C\left( s\right) $ might be calculated in flat
(Minkowski) space-time from the properties of quantum fields in vacuum, but
making a calculation which involves all known fields would be a formidable
task. Nevertheless the calculation is straightforward in principle, as shown
by the derivation which follows of the contribution due to the free
electromagnetic field, which I will perform for illustrative purposes. In
quantum theory the vacuum expectation of the energy of any unexcited field
is assumed to be zero and this assumption is stated formally by using the
normal ordering of the operators. For instance in the electromagnetic field
we have 
\begin{equation}
\left\langle vac\left| \widehat{\rho }\left( \mathbf{r},t\right) \right|
vac\right\rangle =0,\text{ }\widehat{\rho }\left( \mathbf{r},t\right) \equiv 
\text{\thinspace }:\frac{\widehat{\mathbf{E}}\left( \mathbf{r},t\right) ^{2}+%
\widehat{\mathbf{H}}\left( \mathbf{r},t\right) ^{2}}{8\pi }:\text{ },
\label{3b}
\end{equation}
where normal ordering implies that the energy density operator $\widehat{%
\rho }$ contains only products of creation, $\widehat{a}^{+}$, and
annihilation, $\widehat{a}$, operators of photons of the type $\widehat{a}%
\widehat{a},\widehat{a}^{+}\widehat{a}^{+}$ y $\widehat{a}^{+}\widehat{a},$
all of which give a nil vacuum expectation. Therefore the vacuum expectation
of the energy density is zero as assumed. However the two-point correlation
is not zero because the operator 
\begin{equation}
\widehat{C}\left( \mathbf{r}_{1},t_{1};\mathbf{r}_{2},t_{2}\right) \equiv (:%
\frac{\widehat{\mathbf{E}}\left( \mathbf{r}_{1},t_{1}\right) ^{2}+\widehat{%
\mathbf{H}}\left( \mathbf{r}_{1},t_{1}\right) ^{2}}{8\pi }:)(:\frac{\widehat{%
\mathbf{E}}\left( \mathbf{r}_{2},t_{2}\right) ^{2}+\widehat{\mathbf{H}}%
\left( \mathbf{r}_{2},t_{2}\right) ^{2}}{8\pi }:),  \label{3d}
\end{equation}
contains terms of the type $\widehat{a}\widehat{a}\widehat{a}^{+}\widehat{a}%
^{+}$ whose vacuum expectation is finite.

The two-point correlation function of the free electromagnetic field will be
just the vacuum expectation of eq.$\left( \ref{3d}\right) $. The calculation
is straightforward using the plane-waves expansions 
\begin{eqnarray}
\widehat{E}\left( \mathbf{r,}t\right) &=&\sum_{\mathbf{k\varepsilon }}\left( 
\frac{
\rlap{\protect\rule[1.1ex]{.325em}{.1ex}}h%
\omega }{2V}\right) ^{1/2}\left[ \widehat{a}_{\mathbf{k\varepsilon }}\mathbf{%
\varepsilon }\left( \mathbf{k}\right) \exp (i\mathbf{k.r-}i\omega t)+%
\widehat{a}_{\mathbf{k\varepsilon }}^{+}\mathbf{\varepsilon }\left( \mathbf{k%
}\right) \exp (-i\mathbf{k.r+}i\omega t)\right] ,  \nonumber \\
\widehat{H}\left( \mathbf{r,}t\right) &=&\sum_{\mathbf{k\varepsilon }}\left( 
\frac{
\rlap{\protect\rule[1.1ex]{.325em}{.1ex}}h%
}{2V\omega }\right) ^{1/2}[\widehat{a}_{\mathbf{k\varepsilon }}\left( i%
\mathbf{k\times \varepsilon }\left( \mathbf{k}\right) \right) \exp (i\mathbf{%
k.r-}i\omega t)  \nonumber \\
&&-\widehat{a}_{\mathbf{k\varepsilon }}^{+}\left( i\mathbf{k\times
\varepsilon }\left( \mathbf{k}\right) \right) \exp (-i\mathbf{k.r+}i\omega
t)],  \label{4b}
\end{eqnarray}
with standard notation ($V$ is a normalization volume and $\omega \equiv
\left| \mathbf{k}\right| $). The details may be seen in the Appendix and the
result is 
\begin{equation}
C(s\mathbf{)}=\frac{2
\rlap{\protect\rule[1.1ex]{.325em}{.1ex}}h%
^{2}}{\pi ^{4}s^{8}},s^{2}\equiv r^{2}-t^{2}.  \label{4c}
\end{equation}
The two-point correlation depends on distance and time interval only via $%
r^{2}-t^{2}$ as it should. Furthermore it depends on $\left|
r^{2}-t^{2}\right| ,$ so making no distinction between time-like and
space-like intervals.

The correlation $C\left( s\right) ,$ eq.$\left( \ref{4c}\right) ,$ decreases
rapidly at large $\left| r^{2}-t^{2}\right| $ but has a strong divergence
when $r^{2}\rightarrow t^{2}.$ It is plausible to assume that quantum fields
other than the electromagnetic one will give rise to counterterms which
eliminate the divergence. In particular the sums in $\mathbf{k}$ which
appear in eqs.$\left( \ref{4b}\right) $ have been extended to very large
values of $\left| \mathbf{k}\right| ,$ but this is physically absurd. Indeed
the plane waves expansions eqs.$\left( \ref{4b}\right) $ correspond to
assuming an energy $\frac{1}{2}
\rlap{\protect\rule[1.1ex]{.325em}{.1ex}}h%
\omega $ per normal mode of the radiation but, for values of $
\rlap{\protect\rule[1.1ex]{.325em}{.1ex}}h%
\omega $ larger than two electron masses, the electromagnetic field may
produce electron-positron pairs and the study of the radiation field alone
makes no sense. This may be also stated saying that at very high frequencies
the electromagnetic vacuum is polarized, so that the denominators $8\pi $ in
eq.$\left( \ref{3b}\right) $ should be replaced by larger quantities. For a
rigorous treatment we should study the electromagnetic field interacting
with the electron-positron field, but we should also include all other
charged particles, and all particles interacting with these via strong of
weak forces. In summary, a consistent calculation of the correlation
function should involve \textit{all} quantum fields (excluding gravity.) In
the absence of that calculation I shall assume that the tail of the
correlation function $C\left( s\right) $ is given by eq.$\left( \ref{4c}%
\right) $, the photon being the only known massless particle which may exist
freely, but that at small values of $s$ the function $C\left( s\right) $
remains finite. If this is the case, it implies that the contributions of
particles with different masses are not additive, but may cancel each other
to some extent. Another mechanism able to remove the strong divergence of eq.%
$\left( \ref{4c}\right) $ will be discussed at the end of this section. A
simple form to take into account these possibilities is to introduce a
cut-off, substituting the following for eq.$\left( \ref{4c}\right) $%
\[
C(s\mathbf{)}=\frac{2
\rlap{\protect\rule[1.1ex]{.325em}{.1ex}}h%
^{2}}{\pi ^{4}\left( s^{2}+\lambda ^{2}\right) ^{4}},\lambda =\frac{%
\rlap{\protect\rule[1.1ex]{.325em}{.1ex}}h%
}{m}, 
\]
where $m$ is an unknown mass. Putting this into eq.$\left( \ref{eg}\right) $
we get 
\[
\rho _{grav}=-\frac{4}{3\pi ^{3}}\frac{Gm^{6}}{%
\rlap{\protect\rule[1.1ex]{.325em}{.1ex}}h%
^{4}}, 
\]
close to Zeldovich\'{}s proposal eq.$\left( \ref{02}\right) .$ It is not
easy to estimate the value of the cut-off mass $m$ without a detailed
knowledge of the correlation function $C(s)$, eq.$\left( \ref{3a}\right) $,
but we may guess that $m$ is of order the masses of the fundamental
particles of the ``standard model'' rather than of order Planck\'{}s mass.
Thus a model resting upon Zeldovich\'{}s idea predicts a density $\rho _{DE%
\text{ }}$ much closer, if not identical, to eq.$\left( \ref{01}\right) $
than to eq.$\left( \ref{001}\right) .$

The dependence of the two-point correlation on the interval $s$ gives rise
to a paradox which reflects the counterintuitive feature that the
correlation does not decrease with distance. It may be seen as a
straightforward prediction of relativistic theory, but I shall show in the
following that it is really counterintuitive. The paradox may be stated as
follows. Let us assume, for the sake of simplicity, that the density vacuum
fluctuation in a point possesses discrete values, say $\rho _{1}$ with
probability $P_{1},$ $\rho _{2}$ with probability $P_{2},$ etc. Then the
mean square fluctuation will be (compare with eq.$\left( \ref{3a}\right) )$%
\begin{equation}
C\left( 0\right) =\sum_{j}P_{j}\rho _{j}^{2}.  \label{c1}
\end{equation}
Now we consider two different points, the correlation of fluctuations being 
\begin{equation}
C\left( s\right) =\sum_{j}\sum_{k}p_{jk}\rho _{j}\rho _{k},  \label{c2}
\end{equation}
where $p_{jk}$ is the probability that the vacuum fluctuation in the first
point is $\rho _{j}$ and the fluctuation in the second point it is $\rho
_{k} $. If the two said points are separated by a light-like interval, then $%
s=0$ so that eqs.$\left( \ref{c1}\right) $ and $\left( \ref{c2}\right) $
lead to 
\[
C\left( 0\right) =\sum_{j}\sum_{k}p_{jk}\rho _{j}\rho _{k}=\sum_{j}P_{j}\rho
_{j}^{2}=\sum_{j}\sum_{k}p_{jk}\rho _{j}^{2}, 
\]
where the latter equality follows from well known properties of the
probabilities. Hence it is trivial to derive the equality 
\[
\sum_{j}\sum_{k}p_{jk}\left( \rho _{j}^{2}+\rho _{k}^{2}-2\rho _{j}\rho
_{k}\right) =\sum_{j}\sum_{k}p_{jk}\left( \rho _{j}-\rho _{k}\right) ^{2}=0. 
\]
For (positive) probabilities this equality can be true only if $p_{jk}=0$
for any $j\neq k.$ This means that the fluctuations are strictly correlated
in the whole light cone of every point. Furthermore, for two arbitrary
points, $S_{1}$ and $S_{2}$, it is always possible to find another point $S$
which is light-like separated from each one. In fact, all points in the
intersection of the light cones of $S_{1}$ and $S_{2}$ do the job. As a
consequence for all pairs of points the probabilities $p_{jk}$ are zero for
any $j\neq k$. which implies that vacuum fluctuations are \textit{strictly
correlated at all points in space-time}!. This conclusion, asides from being
highly counterintuitive, contradicts known facts about quantum fluctuations.
A possible solution to the paradox is that correlations between events in
different points of space cannot be written, as in eq.$\left( \ref{c2}%
\right) ,$ using joint probabilities, a well known fact in quantum mechanics
(for instance, it is crucial in the proof of Bell\'{}s theorem\cite{Bell}.)
There is another solution (which does not exclude the former), namely that
Minkowski space is not well defined in quantized general relativity. In
fact, in quantized gravity the metric should be quantized, meaning that the
metric coefficients are operators (see eq.$\left( \ref{15}\right) $ below).
Thus neither the distance nor the time interval between events are well
defined. In other words, given two events of coordinates $\left( \mathbf{r}%
_{1},t_{1}\right) $ and $\left( \mathbf{r}_{2},t_{2}\right) $ there is a
quantum uncertainty about the relativistic interval existing between them.
It is possible to state with confidence that two events are spatially
separated if $\left| \mathbf{r}_{1}-\mathbf{r}_{2}\right| >>\left|
t_{1}-t_{2}\right| $ or temporally separated if $\left| \mathbf{r}_{1}-%
\mathbf{r}_{2}\right| <<\left| t_{1}-t_{2}\right| $ , but it is never
possible to state that they are light-like separated. I point out that this
fact already removes the divergence of the two-point correlation function,
shown e. g. in eq.$\left( \ref{4c}\right) .$

\section{Space-time curvature due to quantum vacuum fluctuations}

Working within quantized gravity the space-time structure is determined by
the quantum state, $\mid \Phi \rangle ,$ of the universe and the matter
stress-energy tensor operator, $\widehat{T}_{\mu \nu }\left( x\right) ,$ of
the quantum fields at every space-time point, $x$. Here $x$ stands for the 4
coordinates in an appropriate reference frame, that is 
\begin{equation}
x\equiv \left\{ x_{1},x_{2},x_{3},x_{4}\right\} ,  \label{2.3}
\end{equation}
The study of the quantum fields in curved space-times and the gravitational
back reaction of the fields is a difficult subject\cite{Wald}. In particular
the curvature may give rise to a modification of the vacuum stress-energy%
\cite{ES}. However for our purposes the metric is so close to Minkowskian
that we may treat the quantum fields as if they existed in flat space-time,
although we want to calculate the (small) curvature induced by the vacuum
fluctuations of the fields.

Our approach rests upon the existence of two quite different scales in the
problem, namely a \textit{cosmic} scale (with typical distances of
megaparsecs) and the \textit{atomic }scale of the correlations between
vacuum fluctuations (which involves distances smaller than, say,
nanometers). In the latter scale quantization is essential, but in the
former we may treat everything as classical, as is explained in the
following. For any two quantum observables, $\widehat{a}\left( x\right) $
and $\widehat{b}\left( y\right) $, at the space-time points $x$ and $y$
respectively, we may define the correlation 
\begin{equation}
C_{ab}\left( x,y\right) \equiv \langle \Phi \mid \widehat{a}\left( x\right) 
\widehat{b}\left( y\right) \mid \Phi \rangle -\langle \Phi \mid \widehat{a}%
\left( x\right) \mid \Phi \rangle \langle \Phi \mid \widehat{b}\left(
y\right) \mid \Phi \rangle .  \label{2.4}
\end{equation}
Now it is plausible to assume that the correlation may be relevant at the
atomic scale but goes to zero when the distance increases toward a
macroscopic scale. As a consequence, in the cosmic scale we may treat the
expectations of quantum observables as classical variables, and the
expectations of products of observables as products of the corresponding
classical variables. For instance 
\begin{equation}
\langle \Phi \mid \widehat{a}\left( x\right) \widehat{b}\left( y\right) \mid
\Phi \rangle \simeq a(x)b(y),\text{ }a(x)\equiv \langle \Phi \mid \widehat{a}%
\left( x\right) \mid \Phi \rangle ,\text{ }b(y)\equiv \langle \Phi \mid 
\widehat{b}\left( y\right) \mid \Phi \rangle .  \label{2.5}
\end{equation}
In summary, we may ignore quantization when working with problems at any
macroscopic scale provided we use as classical variables the expectations of
the corresponding quantum observables\textit{. }In sharp contrast, at the
atomic scale we should work within quantized gravity. This happens in
particular when $x=y$, that is 
\[
\langle \Phi \mid \widehat{a}\left( x\right) \widehat{b}\left( x\right) \mid
\Phi \rangle \neq \langle \Phi \mid \widehat{a}\left( x\right) \mid \Phi
\rangle \times \langle \Phi \mid \widehat{b}\left( x\right) \mid \Phi
\rangle . 
\]

The main hypothesis of this paper is that the expectation of the
stress-energy tensor operator of the quantum fields at any point gives the
matter (baryonic or dark) stress-energy, without any additional contribution
of the vacuum. With reference to eqs.$\left( \ref{2.00}\right) $ and $\left( 
\ref{2.2}\right) $, this means that

\begin{equation}
\langle \Phi \mid \widehat{T}_{0}^{0}\mid \Phi \rangle =\rho _{mat},\text{ }%
\langle \Phi \mid \widehat{T}_{\mu }^{\nu }\mid \Phi \rangle \simeq 0\text{
for }\mu \nu \neq 00.  \label{2.6}
\end{equation}
This suggests defining a vacuum stress-energy tensor operator as 
\begin{equation}
\widehat{T}_{\mu \nu }^{vac}\equiv \widehat{T}_{\mu \nu }-\langle \Phi \mid 
\widehat{T}_{\mu \nu }\mid \Phi \rangle \widehat{I}\equiv \widehat{T}_{\mu
\nu }-T_{\mu \nu }^{mat}\widehat{I}  \label{2.7}
\end{equation}
where $\widehat{I}$ is the identity operator. The existence of vacuum
fluctuations means that, although the expectation of $\widehat{T}_{\mu \nu
}^{vac}$ is zero by definition, there are correlated vacuum fluctuations,
that is

\begin{equation}
\left\langle \Phi \left| \widehat{T}_{\mu \nu }^{vac}\left( x\right) 
\widehat{T}_{\lambda \sigma }^{vac}\left( y\right) \right| \Phi
\right\rangle \neq 0\text{ in general.}  \label{2.8}
\end{equation}

In order to proceed with the calculation I shall start with the quantum
metric 
\begin{equation}
d\widehat{s}^{2}=\widehat{g}_{\mu \nu }dx^{\mu }dx^{\nu },  \label{15}
\end{equation}
using polar coordinates 
\begin{equation}
x^{0}=t,\text{ }x^{1}=r,\text{ }x^{2}=\theta ,\text{ }x^{3}=\phi ,
\label{16}
\end{equation}
and I shall write the (quantum operators) coefficients of the metric in the
form 
\begin{eqnarray}
\widehat{g}_{00} &=&-1+\widehat{h}_{00},\;\widehat{g}_{11}=1+\widehat{h}%
_{11},\;\widehat{g}_{22}=r^{2}\left( 1+\widehat{h}_{22}\right) ,  \nonumber
\\
\widehat{g}_{33} &=&r^{2}\sin ^{2}\theta \left( 1+\widehat{h}_{33}\right) ,\,%
\widehat{g}_{\mu \nu }=\widehat{h}_{\mu \nu }\text{ for }\mu \neq \nu ,
\label{17}
\end{eqnarray}
(multiplication of every term times the unit operator is implicit). Here $%
\widehat{h}_{\mu \nu }$ is a (small in some sense) correction to a Minkowski
metric. If we want that the vacuum expectation of eq.$\left( \ref{15}\right) 
$ agrees with eqs.$\left( \ref{2.00}\right) $ to $\left( \ref{rode}\right) $
we should have, to order $O\left( r^{2}\right) $, 
\begin{eqnarray}
\text{ }\left\langle \widehat{h}_{\mu \nu }\right\rangle &=&0\text{ except }%
\left\langle \widehat{h}_{00}\right\rangle =\frac{8\pi G}{3}\left( \rho
_{DE}+\rho _{mat}\right) r^{2},\text{ }  \nonumber \\
\left\langle \widehat{h}_{11}\right\rangle &=&\frac{8\pi G}{3}\left( \rho
_{DE}-\frac{1}{2}\rho _{mat}\right) r^{2},  \label{h}
\end{eqnarray}
where $\left\langle \widehat{h}_{\mu \nu }\right\rangle $ stands for $%
\left\langle \Phi \left| \widehat{h}_{\mu \nu }\right| \Phi \right\rangle .$

The proof will consist of the following steps:

1. We should define an Einstein quantum tensor operator $\widehat{G}_{\mu
\nu }$ in terms of the operators $\widehat{g}_{\mu \nu }$ (or what is
equivalent, the operators $\widehat{h}_{\mu \nu }).$

2. Assuming that in quantized gravity the counterpart of Einstein equations
reads 
\begin{equation}
\widehat{G}_{\mu \nu }=\frac{8\pi G}{c^{4}}\widehat{T}_{\mu \nu },  \label{E}
\end{equation}
we should solve these (non-linear coupled partial differential) operator
equations in order to get the quantum metric coefficients $\widehat{g}_{\mu
\nu }$ in terms of integrals involving the stress-energy tensor operators $%
\widehat{T}_{\mu }^{\nu }\left( x\right) $ and products like $\widehat{T}%
_{\mu }^{\nu }\left( x\right) \widehat{T}_{\lambda }^{\sigma }\left(
y\right) ,$ $\widehat{T}_{\mu }^{\nu }\left( x\right) \widehat{T}_{\lambda
}^{\sigma }\left( y\right) \widehat{T}_{\rho }^{\tau }\left( z\right) ,$ etc.

3. Finally we should calculate the expectation of the metric coefficients $%
\widehat{g}_{\mu \nu }$ in terms of integrals involving the expectations 
\[
\left\langle \Phi \left| \widehat{T}_{\mu }^{\nu }\left( x\right) \right|
\Phi \right\rangle ,\left\langle \Phi \left| \widehat{T}_{\mu }^{\nu }\left(
x\right) \widehat{T}_{\lambda }^{\sigma }\left( y\right) \right| \Phi
\right\rangle ,\left\langle \Phi \left| \widehat{T}_{\mu }^{\nu }\left(
x\right) \widehat{T}_{\lambda }^{\sigma }\left( y\right) \widehat{T}_{\rho
}^{\tau }\left( z\right) \right| \Phi \right\rangle ,etc. 
\]
The expectation of the metric should reproduce eqs.$\left( \ref{h}\right) .$

A problem appears in the first step because there is not yet a quantum
gravity theory specifying $\widehat{G}_{\mu \nu }$ in terms of $\widehat{g}%
_{\mu \nu },$ which would involve a quantum counterpart of Riemann\'{}s
theory. I will not solve the problem in general, but for the approximate
expression of $\widehat{G}_{\mu \nu }$ containing only terms linear or
quadratic in the (small) operators $\widehat{h}_{\mu \nu },$ I will make the
plausible assumption that \textit{the expression of }$\widehat{G}_{\mu \nu }$%
\textit{, in terms of }$\widehat{h}_{\mu \nu },$\textit{\ and their
derivatives with respect to the coordinates, is the same as the one for the
corresponding classical quantities with the rule that the operators should
appear in symmetrical order}. The latter assumption means that the operator
corresponding to the classical product $ab$ will be the quantum expression $%
\frac{1}{2}\left( \hat{a}\widehat{b}+\widehat{b}\hat{a}\right) $.

\section{Quantum Einstein equation and its solution}

In order to simplify de calculations I will introduce the approximation of
retaining, in the expression of $\widehat{G}_{\mu \nu },$ only terms of
zeroth and first order in $\widehat{h}_{\mu \nu },$ except for both $%
\widehat{h}_{00}$ and $\widehat{h}_{11}$, which will be mantained up to
second order. With these approximations our calculation simplifies
substantially by the following reasons. Firstly it may be realized that
terms of zeroth order will not contribute to $\widehat{G}_{\mu \nu }(x)$
because to zeroth order the metric eq.$\left( \ref{17}\right) $ is
Minkowskian. In addition, the terms linear in $\widehat{h}_{\mu \nu }$ with $%
\mu \nu \neq 00$ and $\mu \nu \neq 11$ (and of zeroth order in both $%
\widehat{h}_{00}\ $and $\widehat{h}_{11})$ will not contribute to the
expectations $\left\langle \widehat{G}_{\mu \nu }(x)\right\rangle $ and $%
\left\langle \widehat{G}_{\mu \nu }(x)\widehat{G}_{\lambda \sigma
}(y)\right\rangle $ when eqs.$\left( \ref{h}\right) $ are taken into
account. Consequently we may ignore such terms from now on, which in
practice is equivalent to putting $\widehat{h}_{\mu \nu }=0$ whenever $\mu
\nu \neq 00$ and $\mu \nu \neq 11$. This amounts to replacing the metric eq.$%
\left( \ref{17}\right) $ by 
\begin{equation}
d\widehat{s}^{2}=\exp \left( \widehat{\alpha }\right) dr^{2}+(r^{2}d\theta
^{2}+r^{2}\sin ^{2}\theta d\phi ^{2})\widehat{I}-\exp \left( \;\widehat{%
\beta }\right) dt^{2},  \label{17b}
\end{equation}
where $\widehat{I\text{ }}$ is the identity operators and I have introduced
the new functions 
\[
\widehat{\alpha }\equiv \log \left( 1+\widehat{h}_{11}\right) ,\;\widehat{%
\beta }\equiv \log \left( 1-\widehat{h}_{00}\right) , 
\]
for latter convenience. Eq.$\left( \ref{17b}\right) $ looks like the metric
of a space-time with spherical symmetry in standard coordinates. However
there are two important differences. Firstly the metric tensor is a quantum
operator rather than a classical (c-number) tensor. Secondly the quantities $%
\widehat{\alpha }$ and $\widehat{\beta }$ depend on the coordinates $\theta $
and $\phi $ in addition to the dependence on $t$ and $r,$ typical of
spherical symmetry.

The quantized metric eq.$\left( \ref{17b}\right) $ should be used when
working at the atomic scale, but at the cosmic scale we may use a metric
obtained by the expectation of the former, that is 
\begin{eqnarray}
ds^{2} &=&\langle \Phi \mid d\widehat{s}^{2}\mid \Phi \rangle  \label{2.11}
\\
&=&\langle \Phi \mid \exp \left( \widehat{\alpha }\right) \mid \Phi \rangle
dr^{2}+r^{2}d\theta ^{2}+r^{2}\sin ^{2}\theta d\phi ^{2}-\langle \Phi \mid
\exp (\;\widehat{\beta })\mid \Phi \rangle dt^{2}.  \nonumber
\end{eqnarray}
In terms of $\widehat{\alpha }$ and $\widehat{\beta }$ eqs.$\left( \ref{h}%
\right) $ should be written as follows 
\begin{equation}
\frac{8\pi G}{3}\left( \rho _{DE}+\rho _{mat}\right) r^{2}\simeq \langle
\Phi \mid \exp \left( \widehat{\alpha }\right) \mid \Phi \rangle -1\simeq
\langle \Phi \mid \widehat{\alpha }+\frac{\widehat{\alpha }^{2}}{2}\mid \Phi
\rangle ,  \label{2.12}
\end{equation}
\begin{equation}
\frac{8\pi G}{3}\left( \frac{1}{2}\rho _{mat}-\rho _{DE}\right) r^{2}\simeq
\langle \Phi \mid \exp \left( \widehat{\beta }\right) \mid \Phi \rangle
-1\simeq \langle \Phi \mid \widehat{\beta }+\frac{\widehat{\beta }^{2}}{2}%
\mid \Phi \rangle .  \label{2.13}
\end{equation}
Our aim now is to justify these two equations as deriving from vacuum
fluctuations. Thus I shall obtain the expectations of the right sides of eqs.%
$\left( \ref{2.12}\right) $ and $\left( \ref{2.13}\right) $ in terms of the
correlations (two-point functions) of the density fluctuations of the vacuum
fields. To do that we should begin getting the appropriate quantum Einstein
equations (involving tensor operators) and solving them.

From the metric eq.$\left( \ref{17b}\right) $ it is straightforward to get
the quantum Einstein equations provided we assume that they are similar to
the classical counterparts, as explained above. Two of them do not contain
time derivatives and they are the only ones to be studied here, that is

\begin{eqnarray}
8\pi G\rho &=&G_{0}^{0}=\frac{\alpha }{r^{2}}-\frac{\alpha ^{2}}{2r^{2}}+%
\frac{1}{r}\frac{\partial \alpha }{\partial r}-\frac{\alpha }{r}\frac{%
\partial \alpha }{\partial r}-\frac{1}{2r^{2}}\cot \theta \frac{\partial
\alpha }{\partial \theta }  \label{40} \\
&&-\frac{1}{2r^{2}}\frac{\partial ^{2}\alpha }{\partial \theta ^{2}}-\frac{1%
}{2r^{2}s^{2}}\frac{\partial ^{2}\alpha }{\partial \phi ^{2}}-\frac{1}{4r^{2}%
}\left( \frac{\partial \alpha }{\partial \theta }\right) ^{2}-\frac{1}{%
4r^{2}s^{2}}\left( \frac{\partial \alpha }{\partial \phi }\right) ^{2}, 
\nonumber
\end{eqnarray}
\begin{eqnarray}
-8\pi Gp &=&G_{1}^{1}=\frac{\alpha }{r^{2}}-\frac{\alpha ^{2}}{2r^{2}}-\frac{%
1}{r}\frac{\partial \beta }{\partial r}+\frac{\alpha }{r}\frac{\partial
\beta }{\partial r}-\frac{1}{2r^{2}}\cot \theta \frac{\partial \beta }{%
\partial \theta }  \label{40a} \\
&&-\frac{1}{2r^{2}}\frac{\partial ^{2}\beta }{\partial \theta ^{2}}-\frac{1}{%
2r^{2}s^{2}}\frac{\partial ^{2}\beta }{\partial \phi ^{2}}-\frac{1}{4r^{2}}%
\left( \frac{\partial \beta }{\partial \theta }\right) ^{2}-\frac{1}{%
4r^{2}s^{2}}\left( \frac{\partial \beta }{\partial \phi }\right) ^{2}. 
\nonumber
\end{eqnarray}
Here I have removed the carets of the operators for notational simplicity
which will be also made from now on. But I remember that both $\alpha $, $%
\beta $ and their derivatives are quantum operators and that whenever we
have a product of two of them it is understood that it means symmetrically
orderer product. For instance 
\[
\frac{\alpha }{r}\frac{\partial \beta }{\partial r}\text{ actually means }%
\frac{1}{2}\left( \frac{\widehat{\alpha }}{r}\frac{\partial \widehat{\beta }%
}{\partial r}+\frac{\partial \widehat{\beta }}{\partial r}\frac{\widehat{%
\alpha }}{r}\right) . 
\]
After some algebra eqs.$\left( \ref{40}\right) $ and $\left( \ref{40a}%
\right) $ may be rewritten, in more compact form, 
\begin{equation}
8\pi r^{2}G\rho =\alpha -\frac{\alpha ^{2}}{2}+r\frac{\partial \alpha }{%
\partial r}-r\alpha \frac{\partial \alpha }{\partial r}-\frac{1}{2}\Delta
\alpha +\frac{1}{4}\alpha \Delta \alpha -\frac{1}{8}\Delta (\alpha ^{2}),
\label{2.9}
\end{equation}
\begin{equation}
-8\pi r^{2}Gp=\alpha -\frac{\alpha ^{2}}{2}-r\frac{\partial \beta }{\partial
r}+r\alpha \frac{\partial \beta }{\partial r}-\frac{1}{2}\Delta \beta +\frac{%
1}{4}\beta \Delta \beta -\frac{1}{8}\Delta \left( \beta ^{2}\right) ,
\label{2.10}
\end{equation}
where $\Delta $ is the angular part of the Laplacian operator, that is 
\[
\Delta \equiv \frac{1}{\sin \theta }\frac{\partial }{\partial \theta }\left(
\sin \theta \frac{\partial }{\partial \theta }\right) +\frac{1}{\sin
^{2}\theta }\frac{\partial ^{2}}{\partial \phi ^{2}}. 
\]

I shall start solving the nonlinear partial differential eq.$\left( \ref{2.9}%
\right) ,$ which contains a single unknown, namely the operator $\alpha
\left( r,\theta ,\phi \right) .$ Actually $\alpha $ may also depend on time,
but as eqs.$\left( \ref{40}\right) $ and $\left( \ref{40a}\right) $ do not
contain time derivatives the time $t$ appears as a parameter, rather than
one of the variables of the partial differential equations, and I shall not
write it explicitly. In order to solve eq.$\left( \ref{2.9}\right) $ I will
approximate the solution by a perturbation expansion in powers of the Newton
constant $G$, and work to order $O(G^{2})$, writing the (operator) metric
parameter $\alpha $ in the form 
\begin{equation}
\alpha =G\alpha _{0}+G^{2}\alpha _{1},  \label{3.0}
\end{equation}
If I put this in eq.$\left( \ref{2.9}\right) $ the terms of first order in $%
G $ give the linear (in the unknown $\alpha _{0})$ equation 
\begin{equation}
8\pi r^{2}\rho =\alpha _{0}+r\frac{\partial \alpha _{0}}{\partial r}-\frac{1%
}{2}\Delta \alpha _{0},  \label{3.1}
\end{equation}
whilst the terms of second order give an equation also linear (in $\alpha
_{1})$, namely

\begin{equation}
\alpha _{1}+r\frac{\partial \alpha _{1}}{\partial r}-\frac{1}{2}\Delta
\alpha _{1}=\frac{\alpha _{0}^{2}}{2}+r\alpha _{0}\frac{\partial \alpha _{0}%
}{\partial r}-\frac{1}{4}\alpha _{0}\Delta \alpha _{0}+\frac{1}{8}\Delta
\left( \alpha _{0}^{2}\right) .  \label{3.3}
\end{equation}
The solution of eq.$\left( \ref{3.1}\right) ,$ with the condition that $%
\alpha _{0}=0$ at $r=0$, may be written in simplified notation as 
\begin{equation}
\alpha _{0}=8\pi A(r^{2}\rho ),  \label{3.1a}
\end{equation}
with the meaning 
\begin{equation}
\alpha _{0}\left( \mathbf{r}\right) =8\pi \int_{0}^{r}d^{3}\mathbf{r}%
_{1}A\left( \mathbf{r,r}_{1}\right) \left[ \mathbf{r}_{1}^{2}\rho \left( 
\mathbf{r}_{1}\right) \right] ,\text{ }\mathbf{r}\equiv \left\{ r,\theta
,\phi \right\} ,  \label{3.2}
\end{equation}
where $A$ is a \textit{kernel} to be specified later on. Eq.$\left( \ref{3.3}%
\right) ,$ with the condition that $\alpha _{1}=0$ at $r=0$, may be solved
similarly leading to 
\begin{equation}
\alpha _{1}=A\left[ \frac{\alpha _{0}^{2}}{2}+r\alpha _{0}\frac{\partial
\alpha _{0}}{\partial r}-\frac{1}{4}\alpha _{0}\Delta \alpha _{0}+\frac{1}{8}%
\Delta \left( \alpha _{0}^{2}\right) \right] ,  \label{3.4}
\end{equation}
where $\alpha _{0}$ is given by eq.$\left( \ref{3.1a}\right) .$ Thus the
solution of eq.$\left( \ref{2.9}\right) $ may be written, in simplified
notation (I shall use units $G=1$ from now on, although writing explicitly
Newton\'{}s constant sometimes for the sake of clarity), 
\begin{equation}
\alpha =8\pi A(r^{2}\rho )+64\pi ^{2}A\left[ \frac{1}{2}\left( 1+r\frac{%
\partial }{\partial r}+\frac{1}{4}\Delta \right) [A(r^{2}\rho )]^{2}-\frac{1%
}{4}[A(r^{2}\rho )]\Delta [A(r^{2}\rho )]\right] .  \label{3.5}
\end{equation}

We are interested in the expectations eqs.$\left( \ref{2.12}\right) $ and we
get 
\begin{equation}
\langle \Phi \mid \alpha +\frac{\alpha ^{2}}{2}\mid \Phi \rangle =\langle
\Phi \mid \alpha +\frac{\alpha ^{2}}{2}\mid \Phi \rangle _{mat}+\langle \Phi
\mid \alpha +\frac{\alpha ^{2}}{2}\mid \Phi \rangle _{vac},  \label{3.5a}
\end{equation}
\begin{eqnarray}
\langle \Phi &\mid &\alpha +\frac{\alpha ^{2}}{2}\mid \Phi \rangle
_{mat}\equiv 8\pi A(r^{2}\rho _{mat})+32\pi ^{2}A\left[ \left( 1+r\frac{%
\partial }{\partial r}+\frac{1}{4}\Delta \right) [A(r^{2}\rho
_{mat})]^{2}\right]  \nonumber \\
&&-16\pi ^{2}A\left[ [A(r^{2}\rho _{mat})]\Delta [A(r^{2}\rho
_{mat})]\right] +32\pi ^{2}[A(r^{2}\rho _{mat})]^{2},  \label{3.5b}
\end{eqnarray}
\begin{eqnarray}
\langle \Phi &\mid &\alpha +\frac{\alpha ^{2}}{2}\mid \Phi \rangle
_{vac}\equiv 32\pi ^{2}A\left( 1+r\frac{\partial }{\partial r}+\frac{1}{4}%
\Delta \right) \langle \Phi \left| [A(r^{2}\rho _{vac})]^{2}\right| \Phi
\rangle ,  \nonumber \\
-16\pi ^{2}\langle \Phi &\mid &[A(r^{2}\rho _{vac})]\Delta [A(r^{2}\rho
_{vac})]\mid \Phi \rangle +32\pi ^{2}\langle \Phi \left| [A(r^{2}\rho
_{vac})]^{2}\right| \Phi \rangle ,  \label{3.6}
\end{eqnarray}
where I have taken into account eqs.$\left( \ref{2.6}\right) $ and $\left( 
\ref{2.7}\right) $. The proof is not difficult taking into account that we
work to first order in $\alpha _{1}$ and to second order in $\alpha _{0}$.
Let us consider for instance the term 
\begin{eqnarray*}
\langle \Phi &\mid &\alpha ^{2}\mid \Phi \rangle \simeq \langle \Phi \mid
\alpha _{0}^{2}\mid \Phi \rangle \\
&=&64\pi ^{2}\int d^{3}\mathbf{r}_{1}r_{1}^{2}A\left( \mathbf{r,r}%
_{1}\right) \int d^{3}\mathbf{r}_{2}r_{1}^{2}A\left( \mathbf{r,r}_{1}\right)
\langle \Phi \left| \rho \left( \mathbf{r}_{1}\right) \rho \left( \mathbf{r}%
_{2}\right) \right| \Phi \rangle .
\end{eqnarray*}
Taken eqs.$\left( \ref{2.7}\right) $ and $\left( \ref{2.6}\right) $ into
account the two-point correlation becomes 
\begin{eqnarray*}
\langle \Phi \left| \rho \left( \mathbf{r}_{1}\right) \rho \left( \mathbf{r}%
_{2}\right) \right| \Phi \rangle &=&\langle \Phi \left| \left[ \rho
_{mat}\left( \mathbf{r}_{1}\right) +\rho _{vac}\left( \mathbf{r}_{1}\right)
\right] \left[ \rho _{mat}\left( \mathbf{r}_{2}\right) +\rho _{vac}\left( 
\mathbf{r}_{2}\right) \right] \right| \Phi \rangle \\
&=&\rho _{mat}\left( \mathbf{r}_{1}\right) \rho _{mat}\left( \mathbf{r}%
_{2}\right) +\langle \Phi \left| \rho _{vac}\left( \mathbf{r}_{1}\right)
\rho _{vac}\left( \mathbf{r}_{2}\right) \right| \Phi \rangle .
\end{eqnarray*}
A similar analysis may be made for the other terms. The result is that 
\textit{the expectation }$\langle \Phi \mid \alpha +\alpha ^{2}/2\mid \Phi
\rangle $\textit{\ is the sum of two expressions, one containing only the
matter density, }$\rho _{mat},$\textit{\ and the other one the vacuum
density, }$\rho _{vac},$\textit{\ i. e. there are no cross terms. It may be
also realized that we should solve eq.}$\left( \ref{2.9}\right) $ \textit{at
least up to terms in }$\mathit{G}^{2}$\textit{\ in order to get the leading
term due to the vacuum fluctuations. }This is because the expectation of the
solution to order $G$, eq.$\left( \ref{3.2}\right) ,$ gives no contribution
due to the vanishing of $\left\langle vac\left| \widehat{\rho }\right|
vac\right\rangle ,$ see eq.$\left( \ref{2.6}\right) .$

The terms with $\rho _{mat}$ give the contribution of matter density to the
metric coefficient $\alpha .$ In particular, if we model the matter density
of the universe by a constant it is not difficult to check, taking eq.$%
\left( \ref{c8}\right) $ into account (see below), that the matter term
gives 
\begin{equation}
\langle \Phi \mid \alpha +\frac{1}{2}\alpha ^{2}\mid \Phi \rangle
_{mat}\simeq \frac{2GM}{r}+\frac{2G^{2}M^{2}}{r^{2}},\text{ }M\equiv \frac{%
4\pi }{3}\rho _{mat}r^{3},  \label{3.7}
\end{equation}
which agrees with the second order expansion of $\exp \alpha $ in the well
known Schwarzschild solution 
\[
\exp \alpha =\left( 1-\frac{2GM}{r}\right) ^{-1}. 
\]

\section{Contribution of the vacuum fluctuations to the metric}

In the following I shall calculate the different terms involved in eqs.$%
\left( \ref{3.6}\right) .$ I start solving eq.$\left( \ref{3.1}\right) $ by
writing $\rho \left( r,\theta ,\phi \right) $ and $\alpha _{0}\left(
r,\theta ,\phi \right) $ as expansions in terms of spherical harmonics, that
is 
\begin{eqnarray}
\rho &=&\sum_{lm}\rho _{lm}\left( r\right) Y_{lm}\left( \theta ,\phi \right)
\Rightarrow \text{ }\rho _{lm}\left( r\right) \equiv \int \rho \left(
r,\theta ,\phi \right) Y_{lm}^{*}\left( \theta ,\phi \right) d\Omega ,\text{ 
}  \nonumber \\
\alpha _{0} &=&\sum_{lm}a_{lm}\left( r\right) Y_{lm}\left( \theta ,\phi
\right) \Rightarrow a_{lm}\left( r\right) \equiv \int \alpha _{0}\left(
r,\theta ,\phi \right) Y_{lm}^{*}\left( \theta ,\phi \right) d\Omega .
\label{c5}
\end{eqnarray}
I get from eq.$\left( \ref{3.1}\right) $%
\begin{equation}
8\pi \rho _{lm}r^{2}=\left[ 1+\frac{1}{2}l\left( l+1\right) \right] a_{lm}+r%
\frac{da_{lm}}{dr},  \label{43a}
\end{equation}
whose solution with the initial condition $a_{lm}\left( 0\right) =0$ is 
\begin{equation}
a_{lm}\left( r\right) =8\pi r\int_{0}^{r}(y/r)^{2+l\left( l+1\right) /2}\rho
_{lm}\left( y\right) dy,  \label{c7}
\end{equation}
Taking eqs.$\left( \ref{c5}\right) $ into account I get (see eq.$\left( \ref
{3.2}\right) )$

\begin{equation}
A(\mathbf{r},\mathbf{r}_{\mathbf{1}})\equiv r^{-1}\sum_{lm}Y_{lm}\left(
\theta ,\phi \right) (r_{1}/r)^{l\left( l+1\right) /2}Y_{lm}^{*}\left(
\theta _{1},\phi _{1}\right) .  \label{c8}
\end{equation}

I am interested in the expectations involving $\rho _{vac}$ defined in eq.$%
\left( \ref{3.6}\right) ,$ where I shall write $\left\langle .\right\rangle $
for $\left\langle \Phi \left| .\right| \Phi \right\rangle $ for notational
simplicity. I begin with 
\begin{eqnarray}
\langle \alpha ^{2}\rangle &\simeq &\langle \alpha _{0}^{2}\rangle =\langle
(A[r^{2}\rho ])^{2}\rangle =\int_{0}^{r}dr_{1}\int d\Omega _{1}r_{1}^{2}A(%
\mathbf{r},\mathbf{r}_{\mathbf{1}})  \nonumber \\
&&\times \int_{0}^{r}dr_{2}\int d\Omega _{2}r_{2}^{2}A(\mathbf{r},\mathbf{r}%
_{\mathbf{2}})C(s),  \label{4.1}
\end{eqnarray}
where the two-point correlation function of the density, $C(s),$ was given
in eq.$\left( \ref{3a}\right) $ with 
\begin{equation}
s\equiv \left| \mathbf{r}_{1}-\mathbf{r}_{2}\right| =\sqrt{%
r_{1}^{2}+r_{2}^{2}-2r_{1}r_{2}u},\;u\equiv \cos \theta _{12}.  \label{c11}
\end{equation}
A consequence of the assumption that the correlation $C$ depends only on the
distance $\left| \mathbf{r}_{1}-\mathbf{r}_{2}\right| ,$ rather than on $%
\mathbf{r}_{1}$ and $\mathbf{r}_{2}$ separately, is that the expectation eq.$%
\left( \ref{4.1}\right) $ will not depend on the angular variables $\left(
\theta ,\phi \right) $ and we may average over these variables. Hence it
follows that $l^{\prime }=l,m^{\prime }=m$ and we get 
\begin{eqnarray}
\langle (A[r^{2}\rho ])^{2}\rangle &=&r^{2}\int_{0}^{r}dr_{1}\int d\Omega
_{1}\int_{0}^{r}dr_{2}\int d\Omega _{2}  \nonumber \\
&&\sum_{lm}(r_{1}r_{2}/r^{2})^{l\left( l+1\right) /2+2}Y_{lm}\left( \theta
_{1},\phi _{1}\right) Y_{lm}^{*}\left( \theta _{2},\phi _{2}\right) C(s).
\label{4.2}
\end{eqnarray}
This leads to 
\begin{equation}
\langle (A[r^{2}\rho ])^{2}\rangle =\frac{r^{2}}{4\pi }\int_{0}^{r}dr_{1}%
\int_{0}^{r}dr_{2}\int_{-1}^{1}du\sum_{l}\left( 2l+1\right)
(r_{1}r_{2}/r^{2})^{2+l\left( l+1\right) /2}P_{l}\left( u\right) C(s).
\label{4.4}
\end{equation}
where I have taken into account the following property of spherical harmonic
functions 
\begin{equation}
\sum_{m}Y_{lm}\left( \theta _{1},\phi _{1}\right) Y_{lm}^{*}\left( \theta
_{2},\phi _{2}\right) =\frac{2l+1}{4\pi }P_{l}\left( \cos \theta
_{12}\right) ,  \label{4.3}
\end{equation}
$\theta _{12}$ being the angle between the directions $\left( \theta
_{1},\phi _{1}\right) $ and $\left( \theta _{2},\phi _{2}\right) $ and $%
P_{l}(u)$ a Legendre polynomial. Now I introduce the new variables $\left\{
y,z\right\} $ defined by 
\begin{equation}
r_{1}=y+z/2,\;r_{2}=y-z/2,  \label{4.5}
\end{equation}
substitute an $s$ integration for the $u$ integration and change the order
of the integrals. Thus eq.$\left( \ref{4.4}\right) $ becomes after some
algebra 
\begin{eqnarray}
\langle \alpha ^{2}\rangle &\simeq &\langle (A[r^{2}\rho ])^{2}\rangle
\simeq \frac{1}{4\pi }\int_{0}^{\infty }C(s)sds\sum_{l}\left( 2l+1\right)
\int_{0}^{s}dz\text{ }I_{l}\left( r,z\right)  \label{4.6} \\
&&I_{l}\left( r,z\right) \equiv \int_{z/2}^{r-z/2}dy\left( \frac{%
y^{2}-z^{2}/4}{r^{2}}\right) ^{1+l(l+1)/2}P_{l}\left( 1-\frac{s^{2}-z^{2}}{%
2y^{2}-z^{2}/2}\right) .  \nonumber
\end{eqnarray}

A similar method allows calculating the term $\langle \Phi \left| (A\rho
_{vac})\Delta (A\rho _{vac})\right| \Phi \rangle $. Putting all relevant
terms of eq.$\left( \ref{3.6}\right) $ together we obtain 
\begin{equation}
\left\langle \alpha \right\rangle =8\pi A\left[ \int_{0}^{\infty
}C(s)sds\sum_{l}\left( 2l+1\right) \int_{0}^{s}dz\left( 1+r\frac{\partial }{%
\partial r}+\frac{1}{2}l\left( l+1\right) \right) I_{l}\left( r,z\right)
\right] .  \label{4.6a}
\end{equation}
It may be realized that the main contribution to $\left\langle \alpha
\right\rangle $ comes from high values of $l$. Thus we might neglect $1\ll 
\frac{1}{2}l\left( l+1\right) $, which also shows that the term $\frac{1}{2}%
\alpha _{0}^{2}$ is negligible in comparison with the term $-\frac{1}{4}%
\alpha _{0}\Delta \alpha _{0}$ in eq.$\left( \ref{3.4}\right) .$ Similarly
we may neglect 
\begin{equation}
\langle \frac{1}{2}\alpha ^{2}\rangle \ll \langle \alpha \rangle
\label{4.6c}
\end{equation}
in eq.$\left( \ref{3.6}\right) .$ Proceeding with the calculation of eq.$%
\left( \ref{4.6a}\right) ,$ I get after some algebra 
\begin{eqnarray}
\left( 1+r\frac{\partial }{\partial r}+\frac{1}{2}l\left( l+1\right) \right)
I_{l}\left( r,z\right) &=&\left( \frac{r-z}{r}\right)
^{1+l(l+1)/2}P_{l}\left( 1-\frac{s^{2}-z^{2}}{2r^{2}}\right)  \nonumber \\
&&-\left[ 1+\frac{1}{2}l\left( l+1\right) \right] I_{l}\left( r,z\right) ,
\label{4.6b}
\end{eqnarray}
where I have neglected $z^{2}/2\ll $ $2r^{2}$ in the argument of the
Legendre polynomial. In the quantity $I_{l}\left( r,z\right) $ we may
approximate $2y^{2}-z^{2}/2$ by $2r^{2}$ because only values of $y$ close to 
$r$ contribute substantially to the y-integral. This leads to 
\begin{eqnarray}
I_{l} &\simeq &P_{l}\left( 1-\frac{s^{2}-z^{2}}{2r^{2}}\right)
\int_{z/2}^{r-z/2}dy(y/r)^{2+l(l+1)}\exp \left( -\frac{z^{2}}{8y^{2}}\left[
2+l(l+1)\right] \right)  \nonumber \\
&\simeq &P_{l}\left( 1-\frac{s^{2}-z^{2}}{2r^{2}}\right) \exp \left( -\frac{%
z^{2}}{8r^{2}}\left[ 2+l(l+1)\right] \right)
\int_{z/2}^{r-z/2}dy(y/r)^{2+l(l+1)}  \nonumber \\
&\simeq &r\left[ 3+l\left( l+1\right) \right] ^{-1}P_{l}\left( 1-\frac{%
s^{2}-z^{2}}{2r^{2}}\right)  \nonumber \\
&&\times \exp \left( -\frac{z^{2}}{8r^{2}}\left[ 2+l(l+1)\right] \right)
\left( \frac{r-z/2}{r}\right) ^{2+l(l+1)}  \nonumber \\
&\simeq &r\left[ 3+l\left( l+1\right) \right] ^{-1}P_{l}\left( 1-\frac{%
s^{2}-z^{2}}{2r^{2}}\right) \exp \left( -\frac{z}{2r}\left[ 2+l(l+1)\right]
\right) ,  \label{4.7}
\end{eqnarray}
where I have neglected 
\[
\frac{z^{2}}{8r^{2}}\left[ 2+l(l+1)\right] \ll \frac{z}{2r}\left[
2+l(l+1)\right] 
\]
in the exponential because $z<s\ll r$. Hence eq.$\left( \ref{4.6b}\right) $
becomes 
\begin{eqnarray}
\langle \alpha \rangle &\simeq &4\pi A\ \left[ r\int_{0}^{\infty
}C(s)sds\sum_{l}\left( 2l+1\right) \frac{4+l(l+1)}{3+l(l+1)}\right. 
\nonumber \\
&&\left. \times \int_{0}^{s}dz\exp \left( -\frac{z}{2r}\left[
2+l(l+1)\right] \right) P_{l}\left( 1-\frac{s^{2}-z^{2}}{2r^{2}}\right)
\right] .  \label{4.8}
\end{eqnarray}
We may substitute a Bessel function for the Legendre polynomial, which is a
good approximation for large $l$ and argument close to unity, as is shown by
the well known limit (see Gradshteyn et al.\cite{Grad}, n${{}^{o}}$ 8.722-2) 
\[
\lim_{l\rightarrow \infty }P_{l}\left( \cos \frac{x}{l}\right) =J_{0}\left(
x\right) . 
\]
After that I will substitute an integral for the sum in $l$ and I get

\begin{eqnarray}
\langle \alpha \rangle &\simeq &4\pi A\left[ r\int_{0}^{\infty
}C(s)sds\int_{0}^{s}dz\int_{0}^{\infty }ldl\exp \left[ -zl^{2}/2r\right]
J_{0}\left( \frac{l\sqrt{s^{2}-z^{2}}}{r}\right) \right]  \nonumber \\
&=&4\pi A\left[ r^{2}\int_{0}^{\infty }C(s)sds\int_{0}^{s}\frac{dz}{z}\exp
\left[ \frac{z^{2}-s^{2}}{2zr}\right] \right] ,  \label{4.9}
\end{eqnarray}
where the $l$ integral has been taken from the literature (see Gradshteyn%
\cite{Grad} n${{}^{o}}$ 6.631-4) . The $z$ integral may be performed in
terms of the new variable $t=s^{2}/(2rz),$ which leads to 
\begin{eqnarray}
\int_{0}^{s}\frac{dz}{z}\exp \left[ \frac{z^{2}-s^{2}}{2rz}\right]
&=&\int_{s/2r}^{\infty }\frac{dt}{t}\exp \left[ \frac{s^{2}}{2r^{2}t}%
-t\right] =\int_{s/2r}^{\infty }\frac{dt}{t}\exp \varepsilon \exp \left[
-t\right]  \nonumber \\
&\simeq &\int_{s/2r}^{\infty }\frac{dt}{t}\exp \left[ -t\right] =\log
(2r/s)-0.557,  \label{4.10}
\end{eqnarray}
where I have put $\exp \varepsilon \simeq 1$ because $\varepsilon \equiv
s^{2}/\left( 2r^{2}t\right) \ll 1$ for $t\geq s/2r.$ Now, taking into
account the action of the kernel $A,_{\text{ }}$eq.$\left( \ref{c8}\right) ,$
applied to a function $f\left( \mathbf{r}\right) $ not depending on the
polar angles $\left\{ \theta ,\phi \right\} $, we get finally 
\begin{eqnarray}
\langle \Phi &\mid &\widehat{g}_{11}\mid \Phi \rangle _{vac}\simeq \langle
\Phi \mid \alpha +\frac{\alpha ^{2}}{2}\mid \Phi \rangle _{vac}\simeq
\langle \Phi \mid \alpha \mid \Phi \rangle _{vac}  \label{4.11} \\
&\simeq &2\pi r^{2}\int_{0}^{\infty }C(s)sds\left( \log (2r/s)-0.557\right)
\sim 600r^{2}\int_{0}^{\infty }C(s)sds,  \nonumber
\end{eqnarray}
where I have taken into account eq.$\left( \ref{4.6c}\right) $ and I have
estimated $r/s\sim 10^{40}$. If we include the contribution of matter and
put explicitly Newton\'{}s constant we get 
\begin{equation}
\langle \Phi \mid \alpha +\frac{\alpha ^{2}}{2}\mid \Phi \rangle \simeq 
\frac{8\pi G}{3}\rho _{mat}r^{2}+600G^{2}r^{2}\int_{0}^{\infty }C(s)sds.
\label{4.11d}
\end{equation}

Now I shall solve the second component of the Einstein equation, eq.$\left( 
\ref{2.10}\right) ,$ which involves both coefficients $\alpha $ and $\beta $
of the metric. I will search for a solution of the form 
\[
\beta (r,\theta ,\phi )=-\alpha (r,\theta ,\phi )+\gamma \left( \theta ,\phi
\right) r^{2}+O\left( r^{4}\right) , 
\]
which taking eq.$\left( \ref{2.9}\right) $ into account leads to 
\begin{eqnarray*}
8\pi r^{2}\left( \rho +p\right) &=&2r^{2}\gamma -2r^{2}\alpha \gamma +\frac{1%
}{2}r^{2}\Delta \gamma -\Delta \alpha +\frac{1}{4}\alpha \Delta \alpha \\
&&-\frac{1}{4}(\alpha -\gamma r^{2})\Delta (\alpha -\gamma r^{2})-\frac{1}{8}%
\Delta (\alpha ^{2})+\frac{1}{8}\Delta [(\alpha -\gamma r^{2})^{2}] \\
&=&2r^{2}\gamma -\Delta \alpha +O\left( r^{4}\right) .
\end{eqnarray*}
where in the second equality I have taken into account that $\alpha $ is of
order $O\left( r^{2}\right) $ (see eqs.$\left( \ref{c5}\right) $ and $\left( 
\ref{c7}\right) )$. This leads to the solution 
\[
\beta =-\alpha +4\pi r^{2}\left( \rho +p\right) +\frac{1}{2}\Delta \alpha
+O\left( r^{4}\right) . 
\]
Hence eqs.$\left( \ref{2.12}\right) $ and $\left( \ref{2.13}\right) $ lead
to the following equalities, where terms of order $O\left( r^{4}\right) $
are neglected 
\begin{eqnarray*}
-\langle \Phi \left| \widehat{a}+\frac{\widehat{\alpha }^{2}}{2}\right| \Phi
\rangle &\simeq &-\langle \Phi \mid \widehat{\alpha }\mid \Phi \rangle
\simeq \langle \Phi \mid \widehat{\beta }\mid \Phi \rangle -4\pi r^{2}\rho
_{mat} \\
&\simeq &\langle \Phi \mid \widehat{\beta }+\frac{\widehat{\beta }^{2}}{2}%
\mid \Phi \rangle -4\pi r^{2}\rho _{mat}.
\end{eqnarray*}
The third equality takes into account that $\langle \Phi \mid \widehat{%
\alpha }\mid \Phi \rangle $ does not depend on the angles $\theta ,\phi $,
so that $\langle \Phi \mid \Delta \widehat{\alpha }\mid \Phi \rangle =\Delta
\langle \Phi \mid \widehat{\alpha }\mid \Phi \rangle =0,$ and the last
equality the fact that $\beta $ is of order $O\left( r^{2}\right) .$ Hence,
taking eq.$\left( \ref{4.11d}\right) $ into account, we get 
\begin{equation}
\langle \Phi \mid \beta +\frac{\beta ^{2}}{2}\mid \Phi \rangle \simeq \frac{%
4\pi G}{3}\rho _{mat}r^{2}-600G^{2}r^{2}\int_{0}^{\infty }C(s)sds.
\label{4.11e}
\end{equation}

\section{Conclusions}

The main result of the paper is that eqs.$\left( \ref{4.11d}\right) $ and $%
\left( \ref{4.11e}\right) $ provide the expectation of the quantized metric
(see eq.$\left( \ref{2.11}\right) ),$ in terms of the two-point correlation
of the vacuum fluctuations, $C(s)$, that is 
\begin{eqnarray}
ds^{2} &=&\left( \frac{8\pi G}{3}\rho
_{mat}r^{2}+600G^{2}r^{2}\int_{0}^{\infty }C(s)sds\right)
dr^{2}+r^{2}d\theta ^{2}  \nonumber \\
&&+r^{2}\sin ^{2}\theta d\phi ^{2}-\left( \frac{4\pi G}{3}\rho
_{mat}r^{2}-600G^{2}r^{2}\int_{0}^{\infty }C(s)sds\right) dt^{2}.  \label{5}
\end{eqnarray}
This is identical to the standard FRW metric, eq.$\left( \ref{2.0}\right) ,$
via the approximate ``free falling'' metric eq.$\left( \ref{2.2}\right) ,$
provided that we identify 
\begin{equation}
\rho _{DE}\simeq 140G\int_{0}^{\infty }C(s)sds,  \label{4.11c}
\end{equation}
which shows that vacuum fluctuations give rise to a curvature of space-time
similar to what would be produced by a ``dark energy'' density (plus cold
matter). However we cannot fix the value of $\rho _{DE}$ as far as we do not
know the two-point correlation function of vacuum density fluctuations.
Crucial for the result is the fact that Einstein\'{}s equation involves a 
\textit{non-linear} relation between the stress-energy tensor and the metric
tensor. In fact, if the relation was linear, then the vanishing of the
vacuum expectation of the quantum matter density operator would imply
vanishing of curvature, that is Minkowski space, in the absence of matter.

Our results suggest that the observed accelerated expansion of the universe
might be explained as due to the quantum vacuum fluctuations, without the
need of any ``dark energy''. If this is the case dark energy, $\rho _{DE}$,
appears as a parameter which mimics the effect of the vacuum fluctuations.
The derived relation between the two-point correlation of the fluctuations
and the value of the dark energy parameter would allow calculating the
latter if the said two-point correlation was known. Arguments are given
which suggest that such calculation might give results in agreement with
observations.

\section{Appendix. Two-point correlation of vacuum fluctations of the
radiation field.}

Let us start calculating the two-point correlation 
\begin{equation}
A(\mathbf{r}_{1}\mathbf{,}t_{1;}\mathbf{r}_{2},t_{2})\equiv \left\langle
vac\left| (:\widehat{\mathbf{E}}\left( \mathbf{r}_{1},t_{1}\right) ^{2}:)(:%
\widehat{\mathbf{E}}\left( \mathbf{r}_{2},t_{2}\right) ^{2}:)\right|
vac\right\rangle .  \label{4}
\end{equation}
Taking eqs.$\left( \ref{4b}\right) $ into account it is not difficult to
prove that contributions to $A,$ eq.$\left( \ref{4}\right) ,$ will derive
only from terms with two annihilation operators coming from $(:\widehat{%
\mathbf{E}}\left( \mathbf{r}_{1},t_{1}\right) ^{2}:)$ and two creation
operators coming from $(:\widehat{\mathbf{E}}\left( \mathbf{r}%
_{2},t_{2}\right) ^{2}:).$ Thus the contributing terms will derive from 
\begin{eqnarray*}
&&\frac{
\rlap{\protect\rule[1.1ex]{.325em}{.1ex}}h%
}{V}\sum_{\mathbf{k\varepsilon }}\sum_{\mathbf{k}^{\prime }\mathbf{%
\varepsilon }^{\prime }}\sqrt{\omega \omega ^{\prime }}a_{\mathbf{%
k\varepsilon }}a_{\mathbf{k}^{\prime }\mathbf{\varepsilon }^{\prime }}%
\mathbf{\varepsilon }\left( \mathbf{k}\right) \cdot \mathbf{\varepsilon }%
^{\prime }\left( \mathbf{k}^{\prime }\right) \exp \left[ i\left( \mathbf{k}%
^{\prime }+\mathbf{k}\right) \mathbf{.r}_{1}-i\omega t_{1}\right] \\
&&\times \frac{
\rlap{\protect\rule[1.1ex]{.325em}{.1ex}}h%
}{V}\sum_{\mathbf{k}^{\prime \prime }\mathbf{\varepsilon }^{\prime \prime
}}\sum_{\mathbf{k}^{\prime \prime \prime }\mathbf{\varepsilon }^{\prime
\prime \prime }}\sqrt{\omega ^{\prime \prime }\omega ^{\prime \prime \prime }%
}a_{\mathbf{k}^{\prime \prime }\mathbf{\varepsilon }^{\prime \prime }}^{+}a_{%
\mathbf{k}^{\prime \prime \prime }\mathbf{\varepsilon }^{\prime \prime
\prime }}^{+}\mathbf{\varepsilon }^{\prime \prime }\left( \mathbf{k}^{\prime
\prime }\right) \cdot \mathbf{\varepsilon }^{\prime \prime \prime }\left( 
\mathbf{k}^{\prime \prime \prime }\right) \\
&&\times \exp \left[ -i\left( \mathbf{k}^{\prime \prime }+\mathbf{k}^{\prime
\prime \prime }\right) \mathbf{.r}_{2}+i\omega t_{2}\right] .
\end{eqnarray*}
Taking into account that 
\[
\left\langle vac\left| a_{\mathbf{k\varepsilon }}a_{\mathbf{k}^{\prime }%
\mathbf{\varepsilon }^{\prime }}a_{\mathbf{k}^{\prime \prime }\mathbf{%
\varepsilon }^{\prime \prime }}^{+}a_{\mathbf{k}^{\prime \prime \prime }%
\mathbf{\varepsilon }^{\prime \prime \prime }}^{+}\right| vac\right\rangle
=\delta _{\mathbf{kk}^{\prime \prime }}\delta _{\mathbf{k}^{\prime }\mathbf{k%
}^{\prime \prime \prime }}+\delta _{\mathbf{kk}^{\prime \prime \prime
}}\delta _{\mathbf{k}^{\prime }\mathbf{k}^{\prime \prime }}, 
\]
we get 
\begin{eqnarray}
A &=&\frac{2
\rlap{\protect\rule[1.1ex]{.325em}{.1ex}}h%
^{2}}{V^{2}}\sum_{\mathbf{k\varepsilon }}\sum_{\mathbf{k}^{\prime }\mathbf{%
\varepsilon }^{\prime }}\omega \omega ^{\prime }\left[ \mathbf{\varepsilon }%
\left( \mathbf{k}\right) \cdot \mathbf{\varepsilon }^{\prime }\left( \mathbf{%
k}^{\prime }\right) \right] ^{2}\exp \left[ i\left( \mathbf{k}^{\prime }+%
\mathbf{k}\right) \mathbf{.r-}i\left( \omega +\omega ^{\prime }\right)
t\right]  \nonumber \\
&=&\frac{2
\rlap{\protect\rule[1.1ex]{.325em}{.1ex}}h%
^{2}}{V^{2}}\sum_{\mathbf{k}}\sum_{\mathbf{k}^{\prime }}\omega \omega
^{\prime }\left( 1+\frac{\left( \mathbf{k\cdot k}^{\prime }\right) ^{2}}{%
\mathbf{k}^{2}\mathbf{k}^{\prime 2}}\right) \exp \left[ i\left( \mathbf{k}%
^{\prime }+\mathbf{k}\right) \mathbf{.r-}i\left( \omega +\omega ^{\prime
}\right) t\right] ,  \label{5a}
\end{eqnarray}
where $\mathbf{r=r}_{1}\mathbf{-r}_{2},t\equiv t_{1}-t_{2}$ and I have
performed the sum in polarizations in the second equality. I shall use polar
angles taking the direction of the vector $\mathbf{r}$ as polar axis, so
that 
\[
\mathbf{k\equiv }\omega \left( \sin \theta \cos \phi ,\sin \theta \sin \phi
,\cos \theta \right) ,\mathbf{k\equiv }\omega ^{\prime }\left( \sin \theta
^{\prime }\cos \phi ^{\prime },\sin \theta ^{\prime }\sin \phi ^{\prime
},\cos \theta ^{\prime }\right) . 
\]
After the standard replacements 
\[
\frac{1}{V}\sum_{\mathbf{k}}\rightarrow \frac{1}{8\pi ^{3}}\int_{0}^{\infty
}\omega ^{2}d\omega \int_{-1}^{1}d\left( \cos \theta \right) \int_{0}^{2\pi
}d\phi , 
\]
eq.$\left( \ref{5a}\right) $ gives 
\begin{eqnarray*}
A &=&\frac{
\rlap{\protect\rule[1.1ex]{.325em}{.1ex}}h%
^{2}}{32\pi ^{6}}\int_{0}^{\infty }\omega ^{3}d\omega \int_{0}^{\infty
}\omega ^{\prime 3}d\omega ^{\prime }\int_{-1}^{1}d\left( \cos \theta
^{\prime }\right) \int_{0}^{2\pi }d\phi ^{\prime }\int_{-1}^{1}d\left( \cos
\theta \right) \int_{0}^{2\pi }d\phi \\
&&\times \exp \left[ i\omega ^{\prime }\left( r\cos \theta ^{\prime
}-t\right) \right] \exp \left[ i\omega ^{\prime }\left( r\cos \theta
^{\prime }-t\right) \right] \\
&&\times \left[ 1+\left( \sin \theta \cos \phi \sin \theta ^{\prime }\cos
\phi ^{\prime }+\sin \theta \sin \phi \sin \theta ^{\prime }\sin \phi
^{\prime }+\cos \theta \cos \theta ^{\prime }\right) ^{2}\right] .
\end{eqnarray*}
Integrating the angles $\phi $ and $\phi ^{\prime }$ we get 
\begin{eqnarray*}
A &=&\frac{
\rlap{\protect\rule[1.1ex]{.325em}{.1ex}}h%
^{2}}{16\pi ^{4}}\int_{0}^{\infty }\omega ^{3}d\omega \int_{0}^{\infty
}\omega ^{\prime 3}d\omega ^{\prime }\int_{-1}^{1}du^{\prime
}\int_{-1}^{1}du\left[ 3+3u^{2}u^{\prime 2}-u^{2}-u^{\prime 2}\right] \\
&&\times \exp \left[ i\omega \left( ru^{\prime }-t\right) \right] \exp
\left[ i\omega ^{\prime }\left( ru-t\right) \right] ,\text{ }
\end{eqnarray*}
where $r\equiv \left| \mathbf{r}\right| ,$ $u\equiv \cos \theta ,u^{\prime
}\equiv \cos \theta ^{\prime }.$ The $u$ and $u^{\prime }$ integrals are
trivial and I obtain 
\begin{eqnarray}
A &=&\frac{
\rlap{\protect\rule[1.1ex]{.325em}{.1ex}}h%
^{2}}{16\pi ^{4}}\int_{0}^{\infty }\omega ^{3}d\omega \int_{0}^{\infty
}\omega ^{\prime 3}d\omega ^{\prime }\exp \left[ -i\left( \omega +\omega
^{\prime }\right) t\right]  \nonumber \\
&&\times \left[ 3I\left( x\right) I\left( x^{\prime }\right) +3J\left(
x\right) J\left( x^{\prime }\right) -I\left( x\right) J\left( x^{\prime
}\right) -J\left( x\right) I\left( x^{\prime }\right) \right] ,  \label{6}
\end{eqnarray}
where 
\begin{eqnarray*}
I\left( x\right) &\equiv &\int_{-1}^{1}du\exp \left( -ixu\right) =\frac{%
2\sin x}{x},\text{ }x\equiv \omega r, \\
J\left( x\right) &\equiv &-\frac{d^{2}}{dx^{2}}I\left( x\right) =-\frac{%
2\sin x}{x}-\frac{4\cos x}{x^{2}}+\frac{4\sin x}{x^{3}}.
\end{eqnarray*}
Now 
\begin{eqnarray*}
\int_{0}^{\infty }\omega ^{3}d\omega \exp \left( -i\omega t\right) I\left(
x\right) &=&\frac{2}{r^{4}}\int_{0}^{\infty }x^{2}d\omega \exp \left(
-ixt/r\right) \sin x=\frac{2\left( 3t^{2}+r^{2}\right) }{\left(
r^{2}-t^{2}\right) ^{3}}, \\
\int_{0}^{\infty }\omega ^{3}d\omega \exp \left( -i\omega t\right) J\left(
x\right) &=&-\int_{0}^{\infty }\omega ^{3}d\omega \exp \left( -ixt/r\right) 
\frac{d^{2}}{dx^{2}}\left( \frac{2\sin x}{x}\right) \\
&=&\frac{2\left( t^{2}+3r^{2}\right) }{\left( r^{2}-t^{2}\right) ^{3}}.
\end{eqnarray*}
The correlation between the magnetic energies is the same as eq.$\left( \ref
{4}\right) $ .The correlation between electric and magnetic parts may be
derived without difficulty. It is 
\begin{eqnarray*}
B &=&\frac{2
\rlap{\protect\rule[1.1ex]{.325em}{.1ex}}h%
^{2}}{V^{2}}\sum_{\mathbf{k\varepsilon }}\sum_{\mathbf{k}^{\prime }\mathbf{%
\varepsilon }^{\prime }}\omega \omega ^{\prime }\left[ i\left( \mathbf{k}_{%
\mathbf{r}}\mathbf{\times \varepsilon }\left( \mathbf{k}\right) \right)
\cdot \mathbf{\varepsilon }^{\prime }\left( \mathbf{k}^{\prime }\right)
\right] ^{2}\exp \left[ i\left( \mathbf{k}^{\prime }+\mathbf{k}\right) 
\mathbf{.r-}i\left( \omega +\omega ^{\prime }\right) t\right] , \\
\mathbf{k}_{\mathbf{r}} &=&\frac{\mathbf{k.r}}{\omega r^{2}}\mathbf{r\equiv }%
\omega \left( 0,0,\cos \theta \right) ,
\end{eqnarray*}
whence 
\begin{eqnarray}
B &=&-\frac{2
\rlap{\protect\rule[1.1ex]{.325em}{.1ex}}h%
^{2}}{V^{2}}\sum_{\mathbf{k\varepsilon }}\sum_{\mathbf{k}^{\prime }\mathbf{%
\varepsilon }^{\prime }}\omega \omega ^{\prime }\left( 1+\cos ^{2}\theta
\right) \exp \left[ i\left( \mathbf{k}^{\prime }+\mathbf{k}\right) \mathbf{%
.r-}i\left( \omega +\omega ^{\prime }\right) t\right]  \nonumber \\
&\rightarrow &-\frac{
\rlap{\protect\rule[1.1ex]{.325em}{.1ex}}h%
^{2}}{8\pi ^{4}}\int_{0}^{\infty }\omega ^{3}d\omega \int_{0}^{\infty
}\omega ^{\prime 3}d\omega ^{\prime }  \label{7} \\
&&\times \left[ I\left( x\right) I\left( x^{\prime }\right) +J\left(
x\right) J\left( x^{\prime }\right) \right] \exp \left[ -i\left( \omega
+\omega ^{\prime }\right) t\right] .
\end{eqnarray}
Taking into account eqs.$\left( \ref{6}\right) $ to $\left( \ref{7a}\right) $
we finally obtain 
\begin{equation}
C(r,t\mathbf{)}=2A+B=\frac{2
\rlap{\protect\rule[1.1ex]{.325em}{.1ex}}h%
^{2}}{\pi ^{4}\left( r^{2}-t^{2}\right) ^{4}}.  \label{7a}
\end{equation}


\begin{thebibliography}{99}
\bibitem{Sahni}  Varum Sahni, \textit{Lect. Notes Pys.} \textbf{653},
141-180 (2004).

\bibitem{Hinshaw}  G. Hinshaw et al., \textit{Astrophys. J. Suppl. Ser.} 
\textbf{180}, 225-245 (2009); ArXiv:0803.0732 (astro-ph).

\bibitem{Rich}  J. Rich, \textit{Fundamentals of Cosmology},
Springer-Verlag, Berlin, 2001.

\bibitem{Faraoni}  T. P. Sotiriou, V. Faraoni, \textit{Rev. Mod. Phys. }%
\textbf{82}, 451 (2010)

\bibitem{JCAP}  S. Capozziello, V.F. Cardone, and A. Troisi, \textit{J.
Cosmol. Astropart. Phys.} 0608:001 (2006)

\bibitem{ES}  E. Santos, \textit{Phys. Rev.} \textbf{D} \textbf{81}, 064030
(2010)

\bibitem{Copeland}  E. J. Copeland, M. Sami, S. Tsujikawa, \textit{Int. J.
Mod. Phys. D} \textbf{15}, 1753 (2006); ArXiv:hep-th/0603057.

\bibitem{Weinberg}  S. Weinberg, \textit{Rev. Mod. Phys.} \textbf{61}, 1
(1989).

\bibitem{Zel}  B. Y. Zeldovich, \textit{Sov. Phys. Usp.} \textbf{24}, 216
(1981).

\bibitem{Bell}  J. S. Bell, \textit{Speakable and Unspeakable in Quantum
Mechanics }(Cambridge University Press, Cambridge, 1987). This book contains
reprints of most of Bell\'{}s papers about foundations of quantum mechanics.

\bibitem{Santos}  E. Santos, \textit{Astrophysics Space. Sci.} \textbf{326},
7-10 (2009); \textit{Phys. Lett.} \textbf{A 374}, 709 (2010).

\bibitem{Wald}  R. M. Wald, \textit{Quantum field theory in curved
spacetimes and black hole thermodynamics. }The University of \ Chicago
Press, Chicago (1994).

\bibitem{Grad}  I. S. Gradshteyn, I. M. Ryzhik, A. Jeffrey: \textit{Table of
Integrals, Series, and Products}, Academic Press, Boston (1994).
\end{thebibliography}
\end{document}